# Hemodynamic effects of stent-graft introducer sheath during thoracic endovascular aortic repair

*Yonghui Qiao[1], Le Mao[2], Yan Wang[1], Jingyang Luan[2], Yanlu Chen[1], Ting Zhu[2], Kun Luo[1,]\* and Jianren Fan[1,]\**

[1] *State Key Laboratory of Clean Energy Utilization, Zhejiang University, Hangzhou, China*

[2] *Department of Vascular Surgery, Zhongshan Hospital, Fudan University, Shanghai, China*

\* Corresponding author:

**Kun Luo**

State Key Laboratory of Clean Energy Utilization, Zhejiang University

38 Zheda Road, Hangzhou, China, 310027

Phone: +86-18667160183    E-mail：zjulk@zju.edu.cn

**Jianren Fan**

State Key Laboratory of Clean Energy Utilization, Zhejiang University

38 Zheda Road, Hangzhou, China, 310027

Phone: +86-13336107178    E-mail：fanjr@zju.edu.cn

**Abstract:** Thoracic endovascular aortic repair (TEVAR) has become the standard treatment of a variety of aortic pathologies. The objective of this study is to evaluate the hemodynamic effects of stent-graft introducer sheath during TEVAR. Three idealized representative diseased aortas of aortic aneurysm, coarctation of the aorta, and aortic dissection were designed. Computational fluid dynamics studies were performed in the above idealized aortic geometries. An introducer sheath routinely used in the clinic was virtually-delivered into diseased aortas. Comparative analysis was carried out to evaluate the hemodynamic effects of the introducer sheath. Results show that the blood flow to the supra-aortic branches would increase above 9% due to the obstruction of the introducer sheath. The region exposed to high endothelial cell activation potential (ECAP) expands in the scenarios of coarctation of the aorta and aortic dissection, which indicates that the probability of thrombus formation may increase during TEVAR. The pressure magnitude in peak systole shows an obvious rise and a similar phenomenon is not observed in early diastole. The blood viscosity in the aortic arch and descending aorta is remarkably altered by the introducer sheath. The uneven viscosity distribution confirms the necessity of using non-Newtonian models and high viscosity region with high ECAP further promotes thrombosis. Our results highlight the hemodynamic effects of stent-graft introducer sheath during TEVAR, which may associate with perioperative complications.

# 1. Introduction

The clinical treatment of aortic diseases includes two main methods: open surgery and thoracic endovascular aortic repair (TEVAR). Given the advantages of minimally invasive and low mortality and morbidity, TEVAR has developed to be the clinical standard in the treatment of a variety of common aortic diseases (Czerny et al., 2021). Conventional open surgery continues to play a critical role in late TEVAR failures and complex aortic pathologies, where TEVAR is not suitable to restore the damaged aorta.

During a standard procedure of TEVAR, the stent-graft is delivered to the lesion location by an introducer sheath, which provides the safest method to access the vascular system. However, TEVAR surgery is still met with some serious perioperative events, which are directly related to the success or failure of the operation. Specifically, the perioperative incidence of a stroke may be 4-8%, and spinal cord ischemia occurs in 3-5.6% of patients (Nation and Wang, 2015). Clinical researchers have tried to explain and eliminate perioperative complications. Bismuth et al. (2011) applied transcranial Doppler to detect cerebral embolization during TEVAR and the most dangerous stage could be identified. Uchida (2014) summarized the high-risk factors of spinal cord injury during TEVAR and analyzed how to prevent this event. Air released during implantation of stent-graft could cause stroke and Rylski et al. (2020) evaluated different methods to reduce air volume. Melloni et al. (2021) accessed the outcomes of large-bore introducer sheaths during endovascular aneurysm repair and the application is safe and feasible compared to smaller sheaths.

Recently, image-based *in silico* computational fluid dynamics has provided novel insights into the quantitative hemodynamic analysis of diseased aorta and postoperative aorta. The unstable hemodynamics of aortic dissection and aneurysm was reported by considering the interaction between the blood flow and vessel wall (Alimohammadi et al., 2015; Baumler et al., 2020; Bonfanti et al., 2019; Campobasso et al., 2018; Qiao et al., 2019c). The biomechanics implications of complex TEVAR with the coverage of left subclavian artery, *in-situ* fenestrations, parallel technique, and branched endograft were explored to improve the efficacy of TEVAE (Qiao et al., 2019a; Qiao et al., 2019b; Qiao et al., 2020a; Qiao et al., 2020b; van Bakel et al., 2018; Xu et al., 2019; Zhu et al., 2019). However, the hemodynamic changes during TEVAR have not received enough attention and the potential association with perioperative complications should be clarified.

The present study aims to quantitatively evaluate the hemodynamic effects of stent-graft introducer sheath during TEVAR. In this study, three idealized representative diseased aortas of aortic aneurysm, coarctation of the aorta, and aortic dissection were designed based on the clinical measurement data. An introducer sheath routinely used in the clinic was virtually-delivered into the three diseased aortas. Physiological boundary conditions were coupled to acquire high-precision hemodynamic parameters. The crucial hemodynamics such as blood flow distribution, wall shear stress-related indices, wall pressure, and blood viscosity were comprehensively analyzed to reveal the hemodynamic effects of stent-graft introducer sheath during TEVAR. The founding of this study would help clinicians and manufacturers re-examine the role of introducer sheaths.

## 2. Methodologies

*2.1 Geometry construction and mesh generation*

Following the practice of our previous work (Qiao et al., 2020b), we designed three idealized aortic geometric models of aortic aneurysm, coarctation of the aorta, and aortic dissection, which were generated in SolidWorks (SolidWorks, Waltham, MA). Supra-aortic branches were retained while other relatively small branches of descending aorta were neglected for simplicity. A commonly used stent-graft introducer sheath with a diameter of 14 Fr (4.62 mm) was artificially constructed in the above three diseased aortas to simulate the primary stage of implanting stent-graft during TEVAR. Three-dimensional aortic geometries are shown in Fig. 1.

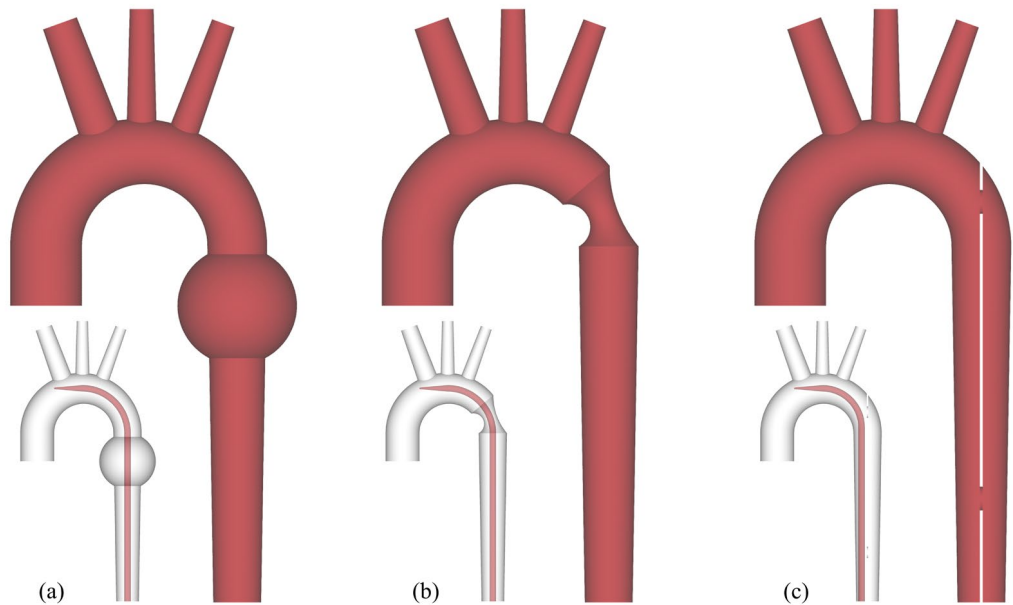

Fig. 1. 3D idealized aortic geometric models. (a) Aortic aneurysm. (b) Coarctation of the aorta. The degree of coarctation is 50%, which is defined as the difference between one and the radius ratio of coarctation to the descending aorta. (c) Aortic dissection. The relatively small figures indicate the comparative scenarios where the stent-graft introducer sheath was virtually added.

Considering the complexity of the diseased aorta, the computational domains of blood flow were discrete into unstructured tetrahedral grids with eight prism boundary layers, which were indispensable to accurately capture the hemodynamic distribution near the aortic wall. Mesh sensitivity analyses were carried out and the chosen meshes all contained more than 2,000,000 grids. The minimal grid size was set to 5E-4 m and the height of the first boundary layer was 1E-4 m.

*2.2 Numerical model and computational details*

Blood was assumed a non-Newtonian incompressible fluid with a density of 1080 kg/m$^3$ and the Carreau-Yasuda model was applied to describe the viscosity change (Gijsen et al., 1999). We neglected the interaction between the blood flow and vessel wall, which was simplified as rigid. A patient-specific pulsatile blood flow waveform (Alastruey et al., 2016) was optimized and coupled in the aortic inlet and a parabolic velocity distribution was adopted to model the injected blood flow from the left heart:

$$u(r)=2u_{ave}(1-\frac{r^2}{R_{inlet}^2}) \qquad (1)$$

where $u_{ave}$ indicates the averaged inlet velocity, $r$ is the radial distance and R denoted the inlet radius. The three-element Windkessel model was coupled in the aortic outlets to predict the pressure waveforms. Parameters of the Windkessel model ($R_p$, $R_d$, *and C*) were acquired by using the workflow proposed by Pirola et al. (2017). For a certain outlet, $R_p$ and $R_d$ were the proximal and distal resistance of the downstream vessel, respectively, and *C* was the compliance. Firstly, we approximated the total resistance ($R_t$), which indicates the sum of $R_p$ and $R_d$ (Les et al., 2010):

$$R_t=\overline{P}/\overline{Q} \qquad (2)$$

where $\overline{P}$ and $\overline{Q}$ are the time-averaged pressure and flow rate. In the present study, a total of 30% blood flow rate was assigned to the three supra-aortic branches, which was redistributed based on the respective outlet area. The aortic systolic and diastolic pressures were 120 and 80 mmHg, respectively and the pressure difference between different outlets was neglected. The average pressure was obtained as (Levick, 2013)

$$\overline{P} = P_{dia} + 1/3(P_{sys} - P_{dia}) \tag{3}$$

The proximal resistance ($R_p$) could be calculated by (Xiao et al., 2014)

$$R_p = \rho c / A \tag{4}$$

where $\rho$ is the blood density and $A$ indicates the outlet area. $c$ is the pulse wave speed (Reymond et al., 2009):

$$c = a_2 / (2r)^{b_2} \tag{5}$$

where $a_2$ (13.3) and $b_2$ (0.3) are constant. $r$ is the outlet radius and it should be emphasized that the unit of the radius is millimeter. The distal resistance ($R_p$) was the difference of the proximal resistance and proximal resistance (LaDisa et al., 2011):

$$R_d = R_t - R_p \tag{6}$$

The compliance ($C$) of the downstream vessel was derived by (Xiao et al., 2014):

$$C = \tau / R_t \tag{7}$$

where $\tau$ (1.79s) is the time constant of the exponential pressure-fall during diastole. The parameters of the Windkessel model from the healthy aorta were used to acquire the blood flow distribution in three diseased computational scenarios. Then we recalculate the specific model parameters based on the above workflow. The same set of model parameters were applied in the corresponding scenario with the introducer sheath. Table 1 shows all the parameters of the Windkessel model in the present study. The time step was 1 ms and all the simulations were carried out on ANSYS Workbench (ANSYS Inc, Canonsburg, USA). The periodic hemodynamic results of the fifth cardiac cycle data were post-processed and present in this study.

Table 1. Parameters of the three-element Windkessel model.

| Scenarios | OUTLET | $R_1$ [$10^7$ Pa s m$^{-3}$] | $C$ [$10^{-10}$ m$^3$ Pa$^{-1}$] | $R_2$ [$10^8$ Pa s m$^{-3}$] |
|---|---|---|---|---|
| Healthy | BT | 6.058 | 21.76 | 7.621 |
| | LCA | 15.49 | 9.620 | 17.06 |
| | LSA | 15.49 | 9.620 | 17.06 |
| | DA | 3.118 | 95.66 | 1.559 |
| Aneurysm | BT | 6.058 | 26.13 | 6.245 |
| | LCA | 15.49 | 10.11 | 16.15 |
| | LSA | 15.49 | 9.648 | 17.00 |
| | DA | 3.118 | 90.77 | 1.660 |
| Coarctation | BT | 6.058 | 26.72 | 6.094 |
| | LCA | 15.49 | 10.56 | 15.40 |
| | LSA | 15.49 | 10.39 | 15.69 |
| | DA | 3.118 | 89.02 | 1.699 |
| Dissection | BT | 6.058 | 29.89 | 5.383 |
| | LCA | 15.49 | 11.40 | 14.16 |
| | LSA | 15.49 | 11.67 | 13.79 |
| | DA | 3.118 | 83.72 | 1.826 |

$R_1$: proximal resistance; $R_2$: distal resistance; $C$: vessel compliance; BT: brachiocephalic artery; LCA: left carotid artery; LSA: left subclavian artery; DA: descending aorta.

## 3. Results

### *3.1 Blood flow distribution*

Adequate blood supply is closely related to the normal physiological function of organs. Fig. 2 shows the comparison of blood flow distribution during a cardiac cycle before and after the delivery of the stent-graft introducer sheath. In the healthy aorta, we assume that 30% incoming blood flow supply to the three supra-aortic branches. Relative to coarctation of the aorta and aortic dissection, aortic aneurysm has minimal impact on the blood flow distribution, where the upward blood flow has increased to 35.59%. For the three aortic diseases, an obvious decrease (9.05~12.98%) is observed on the blood flow supplying to the descending aorta when the introducer sheath is virtually delivered, which may result from the lesion location.

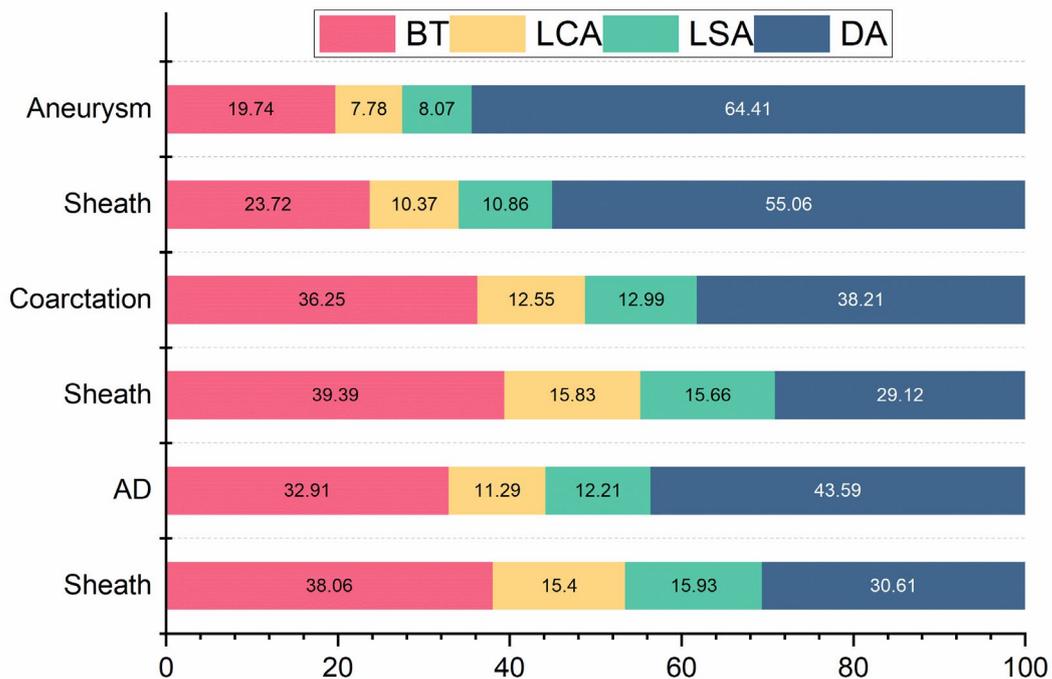

Fig. 2. Comparison of blood flow proportion crossing each outlet before and after the delivery of stent-graft introducer sheath. (BT: brachiocephalic trunk; LCA: left carotid artery; LSA: left subclavian artery; DA: descending aorta.)

*3.2 Wall shear stress-related indices*

The distribution of time-averaged wall shear stress (TAWSS) on the introducer sheath is explored in Fig. 3. The tapered heads of the sheaths are all exposed to high

TAWSS in three aortic diseases. For the coarctation of the aorta, the other high TAWSS region is observed on the sheath wall near the most severe coarctation area. An interesting phenomenon occurs in the aortic dissection scenario. Only the sheath wall near the distal entry of aortic dissection is in a high TAWSS environment, while a similar phenomenon is not observed on the sheath wall near the proximal entry.

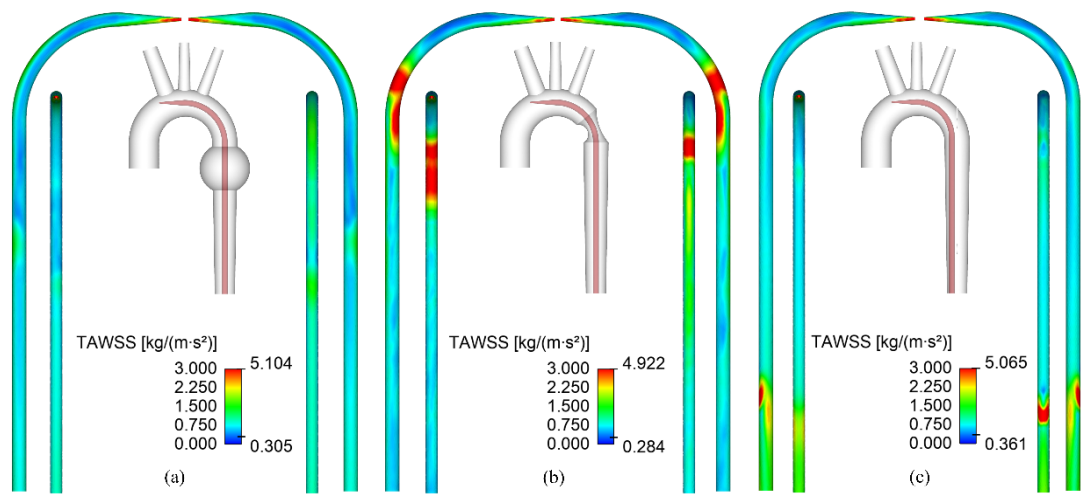

Fig. 3. Distribution of time-averaged wall shear stress (TAWSS) on the stent-graft introducer sheath. (a) Aortic aneurysm. (b) Coarctation of the aorta. (c) Aortic dissection.

We also analyze the distribution of the TAWSS and oscillatory shear index (OSI) on the aortic wall and there is no significant difference before and after the delivery of the stent-graft introducer sheath, therefore it is not shown in the present study. Fig. 4 reveals the distribution of endothelial cell activation potential (ECAP), which could identify the probability of thrombus formation. The wall of the aortic aneurysm shows high ECAP and the introducer sheath has negligible impact on the ECAP distribution. The aortic arch upstream of coarctation and distal coarctation turn to high ECAP when the introducer sheath is present. In aortic dissection, high ECAP is observed on the false lumen except for the region around the proximal entry. After the delivery of the introducer sheath, the aortic arch upstream of proximal entry and false lumen near the distal entry exposed to high ECAP expands remarkably, which indicates that the probability of thrombus formation increases.

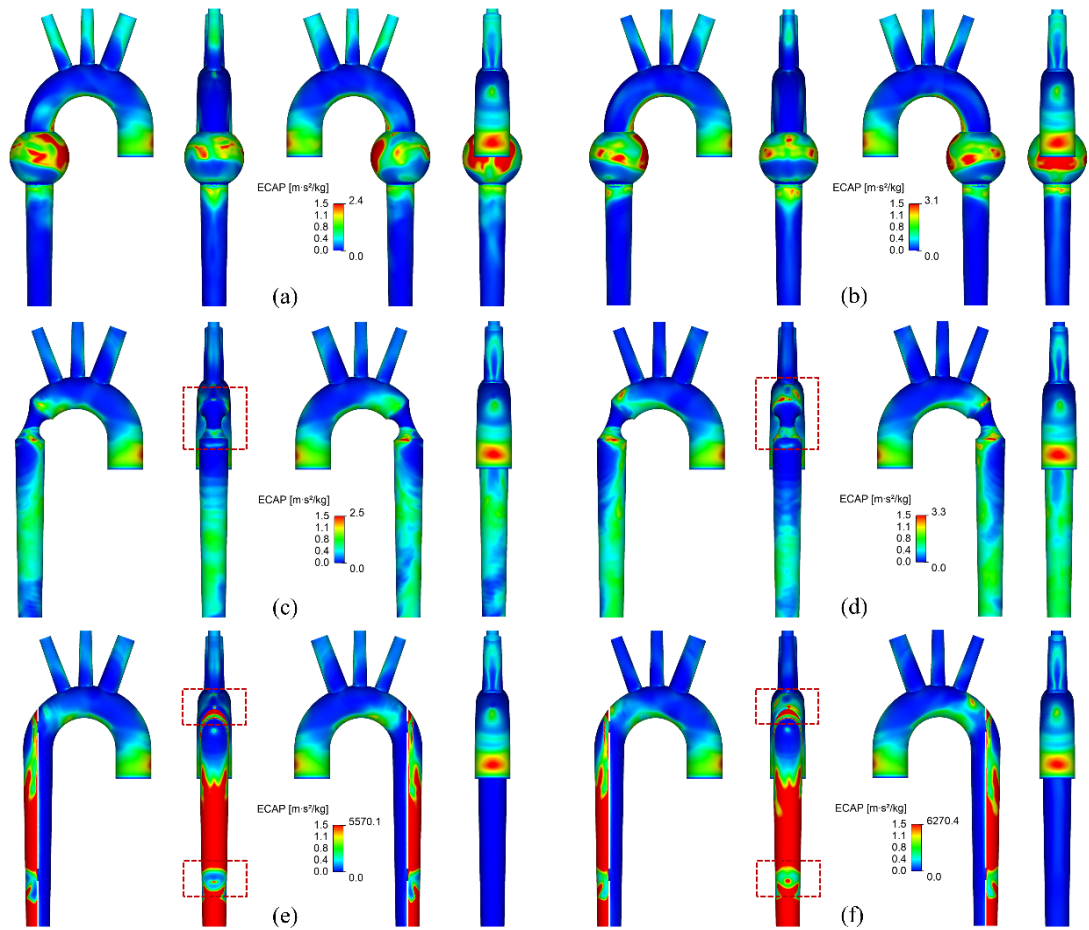

Fig. 4. Comparison of endothelial cell activation potential (ECAP) distribution. (a) Aortic aneurysm. (b) Aortic aneurysm with the introducer sheath. (c) Coarctation of the aorta. (d) Coarctation of the aorta with the introducer sheath. (e) Aortic dissection. (f) Aortic dissection with the introducer sheath.

### *3.4 Wall pressure and blood viscosity*

The distribution of wall pressure in peak systole and early diastole is shown in Fig. 5. The presence of the stent-graft introducer sheath causes a pressure increase on the ascending aorta and aortic arch in peak systole. Specifically, the most remarkable increase is observed in the aortic dissection (9.79%). While there is little change in the wall pressure in early diastole.

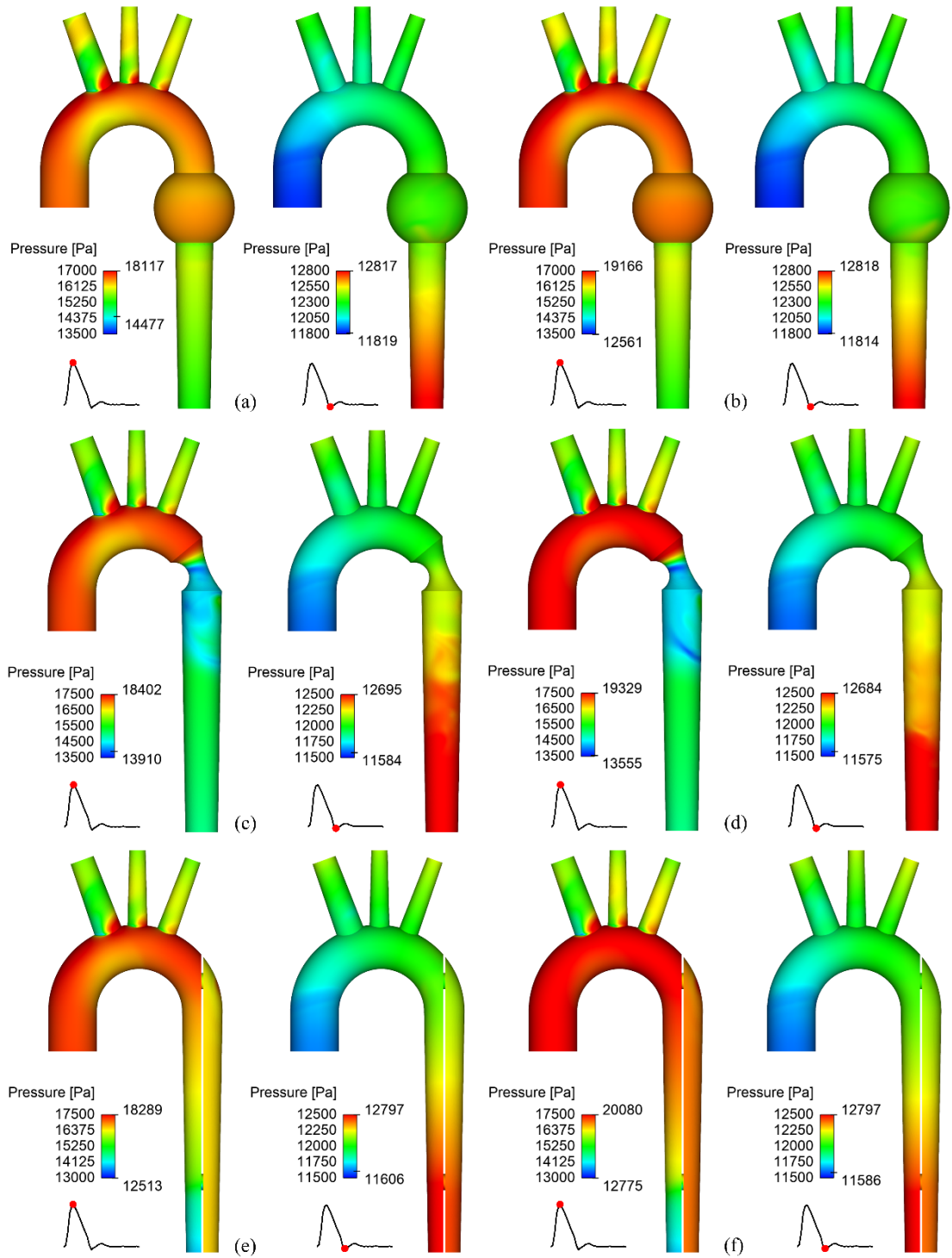

Fig. 5. Comparison of wall pressure in peak systole and early diastole. (a) Aortic aneurysm. (b) Aortic aneurysm with the introducer sheath. (c) Coarctation of the aorta. (d) Coarctation of the aorta with the introducer sheath. (e) Aortic dissection. (f) Aortic dissection with the introducer sheath.

Fig. 6 investigates the interior distribution of time-averaged blood viscosity. In the aortic aneurysm, high viscosity is observed in ascending aorta, aortic branches, and distal descending aorta. The introducer sheath alters the viscosity distribution and causes high viscosity in the lesser curvature of the aortic arch and aneurysm. For coarctation of the aorta, the relatively low viscosity region in the most severe coarctation becomes high viscosity area due to the presence of the introducer sheath. In aortic dissection, the wall of the true lumen shows low viscosity, while high viscosity is observed in the false lumen. After the delivery of the sheath, the low viscosity region in the true lumen disappears and the high viscosity region in the distal false lumen expands.

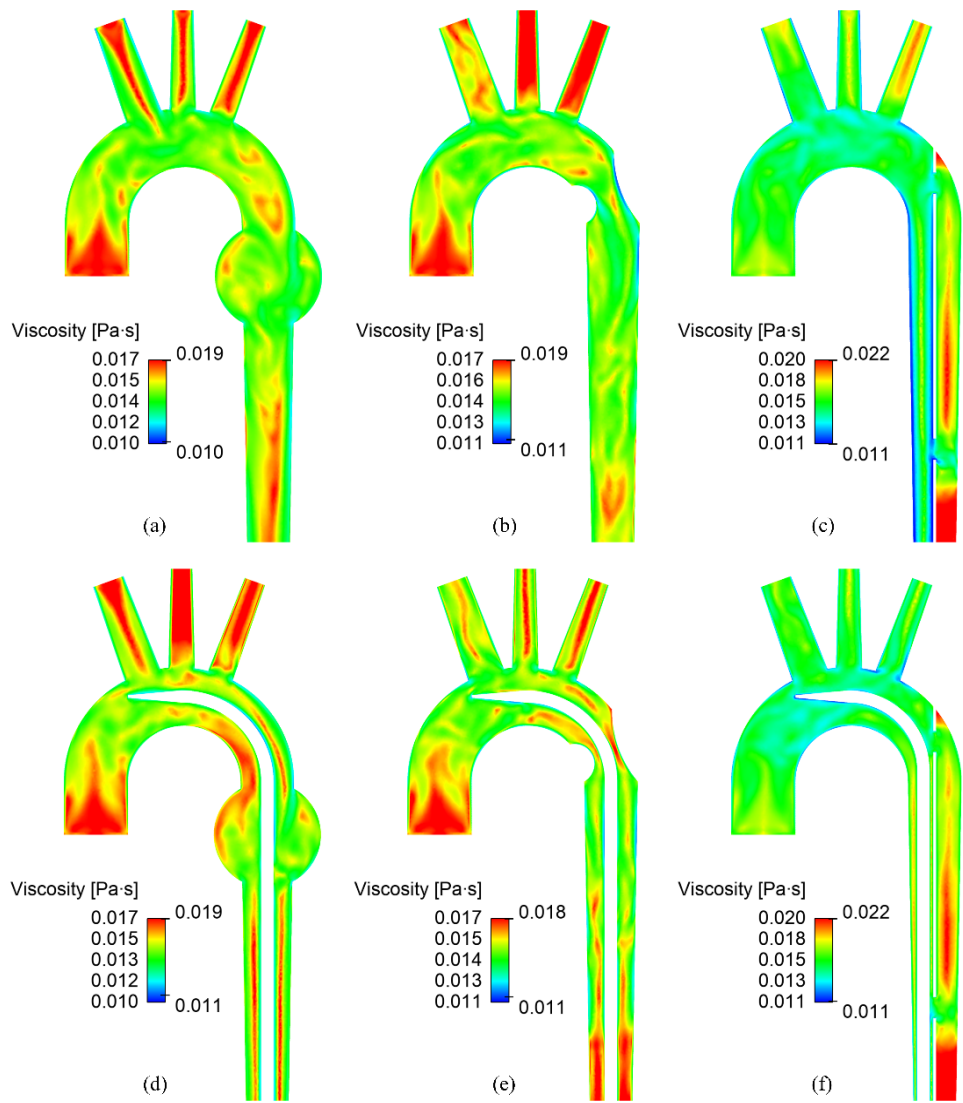

Fig. 6. Comparison of blood viscosity in axial cross-section. (a) Aortic aneurysm. (b) Coarctation of the aorta. (c) Aortic dissection. (d) Aortic aneurysm with the introducer sheath. (e) Coarctation of the aorta with the introducer sheath. (f) Aortic dissection with the introducer sheath.

## 4. Discussion

The perioperative hemodynamic variation may have a potential relationship with severe complications during TEVAR. The present study was designed to evaluate the hemodynamic effects of the stent-graft introducer sheath in idealized aortic aneurysm, coarctation of the aorta, and aortic dissection using the computational fluid dynamics method.

Blood flow to the descending aorta is decreased due to the introducer sheath, which reduces the flow area of descending aorta. The blood that originally supplied the descending aorta turns to the supra-aortic branches. This phenomenon may cause ischemia in the organs connected to the descending aorta. Especially when the implantation of the stent-graft is slow and time-consuming, intraoperative monitoring of related organs becomes particularly necessary.

ECAP distribution is used to characterize the probability of thrombus formation (Kelsey et al., 2017). The high ECAP regions both expand in coarctation of the aorta and aortic dissection. The common location is the aortic arch upstream of the coarctation and proximal entry of the dissection, where the possible thrombus may cause severe perioperative complications. Therefore, we suggest the implantation of the stent-graft should be controlled within a reasonable safe time. In aortic dissection, there is another kind of situation. Results show that the probability of thrombus formation in distal false lumen increases, which is beneficial to the remodeling of the false lumen.

We found that non-Newtonian characteristic of blood flow is obvious in the aorta. Most previous studies have treated blood as a Newtonian fluid, which neglects the response of viscosity to the variation of shear stress rate. Generally, our interest region is the aortic wall, where the high shear stress rate would cause low viscosity. Physiologically relevant wall shear stress is equal to the product of shear rate and viscosity. Therefore, treating the blood viscosity as a constant would overpredict the wall shear stress. It should be emphasized that the high viscosity region coincides with the high ECAP area, which would further increase the probability of thrombus formation in three diseased aortas.

In this paper, the aortic wall was assumed to be rigid, which may fail to identify the low TAWSS and high OSI region in aortic dissection (Alimohammadi et al., 2015). However, our previous study has pointed that the interaction between the blood flow and aortic wall has a negligible impact on the wall shear stress distribution in relatively simple aortic geometries (Qiao et al., 2021). Coarctation of the aorta and aortic aneurysm were considered simple geometric models in the present study. Therefore, we applied rigid assumption to seek a balance between accuracy and computational resource.

## 5. Conclusions

This study evaluates the hemodynamic effects of stent-graft introducer sheath during TEVAR. After the delivery of the sheath, we found the blood flow supplying the supra-aortic branches would increase obviously, which may cause ischemia in the organs connected to the descending aorta. The second major finding is that the presence of the sheath could increase the probability of thrombus formation in distal coarctation and false lumen according to the distribution of ECAP and blood viscosity. Besides, the necessity of adopting a non-Newtonian viscous model is also confirmed. In conclusion, the introducer sheath could alter crucial hemodynamics during TEVAR, which may be associated with perioperative complications.

**Acknowledgment**

This research was supported by the National Postdoctoral Program for Innovative Talents (CN) [grant number BX20200290], Postdoctoral Science Foundation (CN) [grant number 2020M681852], Postdoctoral Science Foundation of Zhejiang Province (CN) [grant number ZJ2020153].

**Conflict of interest**

All authors declare that they have no conflicts of interest.